\documentclass[a4paper]{jpconf}
\usepackage{graphicx}
\begin{document}
\title{Neutrinos and rare isotopes}

\author{A B Balantekin}

\address{Physics Department, University of Wisconsin, Madison WI 53706 USA}

\ead{baha@physics.wisc.edu}
 
\begin{abstract}
The close connection between neutrino physics and the physics explored at rare
isotope facilities is explored. The duality between the Hamiltonian describing the
self-interacting neutrino gas near the proto-neutron star in a core-collapse
supernova and the BCS theory of pairing is elucidated.  This many neutrino system is
unique as it is  the only many-body system driven by weak interactions. Its
symmetries are discussed. 
\end{abstract}


As you read the title of this contribution to the proceedings of the conference
celebrating contributions of Taka Otsuka to nuclear physics on the occasion of his
60th birthday, you may wonder what neutrinos have to do with exotic nuclei. The
answer follows from an examination of the scientific motivations of the physics
enterprise with rare isotopes. Exploring physics with rare isotopes would impact
many areas of inquiry. 
It would contribute to our understanding of nuclear structure by exploring the
limits of nuclear existence, new shapes and new collective behaviors. Studies of
nuclear astrophysics with rare isotope facilities encompass a broad range of
subjects including searches for the origin of elements, investigating rare isotopes
present in explosive phenomena in astrophysical settings, and understanding neutron
star crusts. Rare isotopes can be viewed as laboratories to study fundamental
symmetries  where symmetry violations are sometimes amplified. Last, but not the
least, research at rare isotope facilities would lead to societal applications.
Neutrinos connect all these venues, forming an intellectual bridge between them.

One topic where the connection between rare isotope physics and neutrino physics is
manifest is the nucleosynthesis of various elements. A complete understanding of
element nucleosynthesis requires input from both areas. Not only the knowledge of
neutrino properties such as masses and mixing angles, but also the spin-isospin
response of a broad range of nuclei from stable isotopes to rare ions are the
crucial components of a successful description of the element nucleosynthesis.  This
synergy is perhaps most obvious for the $\nu$-process, the nucleosynthesis via
neutrino-induced nucleon emission \cite{Woosley:1989bd}. The most quoted example of
the $\nu$-process nucleosynthesis is the production of $^{19}$F  by neutrino capture
on $^{20}$Ne in the outer shells of the supernovae,  which accounts for the entire
observed abundance of $^{19}$F. Of course the $\nu$-process in more broadly
operational. For example, recent work nicely ties the $\nu$-process nucleosynthesis
yields of $^{11}$B and $^7$Li in the He shells of supernovae to the neutrino
properties \cite{Yoshida:2005uy}. Matter-enhanced neutrino oscillations governed by
the mixing between the first and the third neutrino generations is operative at
matter densities that exist in those outer shells of a supernova. This boosts the
$\nu$-process nucleosynthesis yields of $^{11}$B and $^7$Li for the inverted
neutrino mass hierarchy, but not for the normal one. 

The cosmic site of the r-process nucleosynthesis,  driven by a succession of rapid
neutron captures on heavy seed nuclei, 
is still unknown \cite{Qian:2007vq},  but the neutrino-driven wind in the
core-collapse supernovae is among the possible sites. 
Recent hydrodynamical simulations of the neutrino-driven wind do not seem to always
result in the necessary extreme conditions \cite{wind}. However, since our
theoretical understanding of core-collapse supernovae is still evolving, it is not
possible to rule out the neutrino-driven wind as a possible site.  Numerical
modeling of core-collapse supernovae has made significant recent progress, in
particular  two- and three-dimensional models were successfully implemented 
unveiling a complex interplay between neutrino physics, thermonuclear reactions and
turbulence \cite{Lentz:2012xc,Mueller:2012is,Brandt:2010xa}. Core-collapse
supernovae are neutrino-dominated dynamical systems 
\cite{Balantekin:2013gqa} 
and one crucial ingredient to understand the r-process nucleosynthesis they may host
 is a better knowledge of neutrino physics. 
A second crucial input is a better working understanding of the spin-isospin
response of a broad range of nuclei.

There is another, more formal, connection between neutrino physics and the nuclear
many-body problem. The sheer number of neutrinos 
($\sim10^{58}$) emitted from the cooling proto-neutron star following the core
collapse necessitates inclusion of neutrino-neutrino interactions in the description
of the neutrino transport in supernovae. Unlike the one-body Hamiltonian of the
matter-enhanced neutrino oscillations where neutrinos interact with a mean-field
(generated by the background particles other than neutrinos), the Hamiltonian
describing the many-neutrino gas in a core-collapse supernova contains both one- and
two-body terms, making it technically much more challenging. Inclusion of the
neutrino-neutrino interaction terms leads to very interesting collective effects
\cite{Duan:2010bg,Raffelt:2012kt}. In contrast to the many-body systems studied in
condensed-matter physics where the systems are primarily driven by electromagnetic
interactions, and in nuclear physics, where the systems are primarily driven by
strong interactions, the many-neutrino system in the core-collapse supernovae is the
only many-body system in nature driven by the weak interactions. In addition, the
latter system contains many more particles than the other two. Partially motivated
by these features many-body aspects of the supernova neutrinos 
recently started to receive more attention
\cite{Balantekin:2006tg,Pehlivan:2011hp,Volpe:2013uxl}. 

There is a further analogy between supernova neutrino many-body system and the
pairing problem in nuclear physics. To demonstrate this  analogy let us consider
only two flavors of neutrinos: electron neutrino, $\nu_e$, and 
another flavor, $\nu_x$ which mixes with the electron neutrinos. For simplicity, we
will also ignore the antineutrinos, which are of course present in
supernovae\footnote{Inclusion of antineutrinos 
requires a second set of SU(2) algebras. For three flavors two sets of SU(3)
algebras are needed.}.  
Introducing the creation and annihilation operators for a neutrino 
with three momentum ${\bf p}$, we can write down the generators of the neutrino
flavor isospin algebras   
\cite{Balantekin:2006tg}: 
\begin{eqnarray}
J_+({\bf p}) &=& a_x^\dagger({\bf p}) a_e({\bf p}), \> \> \>
J_-({\bf p})=a_e^\dagger({\bf p}) a_x({\bf p}), \nonumber \\
J_0({\bf p}) &=& \frac{1}{2}\left(a_x^\dagger({\bf p})a_x({\bf p})-a_e^\dagger({\bf
p})a_e({\bf p})
\right). \label{su2}
\end{eqnarray}
The integrals of these operators over all possible values of momenta generate the
global flavor 
isospin algebra. Using the operators in Eq. (\ref{su2}) 
the Hamiltonian for a neutrino propagating through matter takes the form  
\begin{equation}
\label{total}
H = H_{\nu} + H_{\nu \nu} 
= \left(
\sum_p\frac{\delta m^2}{2p}\hat{B}\cdot\vec{J}_p  - \sqrt{2} G_F 
N_e  J_p^0  \right) 
+ \frac{\sqrt{2}G_{F}}{V}\sum_{\mathbf{p},\mathbf{q}}\left(1- 
\cos\vartheta_{\mathbf{p}\mathbf{q}}\right)\vec{J}_{\mathbf{p}}\cdot\vec{J}_{\mathbf{q}}
\end{equation} 
where the auxiliary vector quantity $\hat{B}$ is given by 
\begin{equation}
\hat{B} = (\sin2\theta,0,-\cos2\theta) ,
\end{equation}
$N_e$ is the background electron density and $\delta m^2$ is the difference between
squares of the masses associated with mass eigenstates. In the above equations
$\theta$ is the mixing angle between electron 
neutrino and the other neutrino flavor and $\vartheta_{\bf pq}$  is the angle
between neutrino momenta ${\bf p}$ and {\bf q}.  
In writing this equation 
a term proportional to identity is omitted as such terms do not contribute to the
neutrino oscillations. 
Note that the presence of the $(1-\cos\vartheta_{\bf pq}) $ term in 
the integral above is crucial to recover the effects of the Standard Model weak  
interaction physics in the most general situation. In the idealized case of an 
isotropic neutrino distribution and a very large number of neutrinos, this term may
average to a 
constant and the neutrino-neutrino interaction term in the 
Hamiltonian simply reduces to the Casimir operator of the global SU(2) algebra. If
the background electron neutrino density is negligible  (a good approximation near
the proto-neutron star where neutrino-neutrino interactions are dominant) and a
single angle dominates the second term, the Hamiltonian of Eq. (\ref{total})  takes
the form 
\begin{equation}
\label{4}
H 
= 
\sum_p\frac{\delta m^2}{2p}\hat{B}\cdot\vec{J}_p
+\frac{\sqrt{2}G_{F}}{V}\vec{J}\cdot\vec{J}, 
\end{equation}
where $\vec{J} = \sum_p \vec{J}_p$. The Hamiltonian in Eq. (\ref{4}) is
mathematically very similar to the BCS Hamiltonian in the quasi-spin
basis\footnote{Note that neutrino Hamiltonian in the single-angle approximation is
{\it not} identical to the BCS Hamiltonian as the sign of the two-body term is
different. Also the Hamiltonian in Eq. (\ref{4}) does not include the antineutrino
terms.}
\begin{equation}
\label{5}
\hat{H}_{\mbox{\tiny
BCS}}=\sum_{k}2\epsilon_{k}\hat{t}_{k}^{0}-|G|\hat{T}^+\hat{T}^- ,
\end{equation}
where the quasi-spin operators also form an SU(2) algebra:
\begin{equation}
\label{6}
[\hat{t}_k^+,\hat{t}_l^-]=2\delta_{kl}\hat{t}_k^0~,\qquad
[\hat{t}_k^0,\hat{t}_l^\pm]=\pm\delta_{kl}\hat{t}_k^\pm~ 
\end{equation}
with $T^+ = \sum_i \hat{t}_i^+$ and so on.  An exact solution of this problem was
given by Richardson some time ago \cite{rich}. This solution was later generalized
by Gaudin \cite{Gaudin1}. A similar solution also exists for the neutrino
Hamiltonian of Eq. (\ref{4}) \cite{Pehlivan:2011hp}. It is manifest that the
Hamiltonian of Eq. (\ref{4}) possesses an $SU(N)_f$ rotation symmetry in the
neutrino flavor space with N flavors
\cite{Balantekin:2006tg,Duan:2008fd,Balantekin:2009dy}. Since the BCS Hamiltonian is
shown to be integrable, there must be  
constants of motion associated with it \cite{yuzb}. One can write down the constants
of motion of the collective neutrino Hamiltonian in Eq. (\ref{4}) as
\cite{Pehlivan:2011hp}
\begin{equation}
\label{7}
\hat{h}_{p} = \hat{B}\cdot\vec{J}_p+2\sum_{q\left(\neq
p\right)}\frac{\vec{J}_{p}\cdot\vec{J}_{q}}{\omega_{p}-\omega_{q}}.
\end{equation} 
where we defined $ \omega_{p}=\frac{1}{\mu}\frac{\delta m^{2}}{2p}$ with
$\mu=\frac{\sqrt{2}G_{F}}{V}$. The Hamiltonian of Eq. (\ref{4}) itself can be
written to include a linear combination of these invariants: 
\begin{equation}
\label{8}
H = \sum_{p}w_{p}\hat{h}_{p}+\sum_{p} \vec{J}_{p}\cdot\vec{J}_{p} 
\end{equation}
It was shown that existence of such constants of motion could lead to collective
neutrino oscillations \cite{Raffelt:2011yb}. 

To solve the full Hamiltonian of Eq. (\ref{total}) for the large number of neutrinos
present in the supernovae is a numerically very challenging, if not almost
impossible, task. The full Hamiltonian is usually simplified using an RPA-like
linearization procedure by writing the two-body terms as a one-body operator times
its expectation value in a suitably chosen basis. For example applying this
procedure to the Hamiltonian of Ref. (\ref{4}) one gets
\begin{equation}
\label{9}
H \rightarrow \hat{H}^{\mbox{\tiny RPA}} =  \sum_p\omega_{p}\hat{B}\cdot\vec{J}_p
+\vec{P}\cdot\vec{J}
\end{equation}
where the polarization vector $\vec{P}$ ensues from the linearization of the flavor
isospin operators: 
\begin{equation} 
\label{10}
\vec{P}_{\mathbf{p},s}=2\langle\vec{J}_{\mathbf{p},s}\rangle . 
\end{equation}
Using the SU(2) coherent states associated with the flavor isospin to calculate the
operator averages 
in the above equations yields a reduced collective neutrino Hamiltonian, widely used
in the 
literature \cite{Balantekin:2006tg}. This introduces an approximation to the
description of the neutrino gas, however abandoning this approximation seems to be
unlikely in the near future for practical reasons. 

Collective neutrino oscillations produce an interesting effect, called spectral
swappings or splits, on the final neutrino energy spectra: at a particular energy
these spectra are almost completely divided into parts of different flavors
\cite{Raffelt:2007cb,Duan:2008za}. 
The nature of the such swaps can be elucidated by referring to the Hamiltonian in
Eq. (\ref{9}). As in the BCS theory, this Hamiltonian does not conserve particle
number. Particle number conservation can be enforced by introducing a Lagrange
multiplier: 
\begin{equation}
\hat{H}^{\mbox{\tiny RPA}} \rightarrow \hat{H}^{\mbox{\tiny RPA}}+\omega_c\hat{J}^0.  
\end{equation}
Diagonalization of this Hamiltonian using an appropriately chosen Bogoliubov
transformation  gives rise to spectral splits or swaps in the neutrino energy
spectra with the Lagrange multiplier playing the role of the swap frequency
\cite{Pehlivan:2011hp}.  

It should be emphasized that the collective neutrino oscillations could
significantly impact r-process nucleosynthesis yields 
\cite{Balantekin:2004ug,Duan:2010af}. 
To understand the r-process is clearly one of the drivers of the experiments with
rare ion beams. This physics program requires measuring the beta-decay rates of
nuclei both at and far from stability, in particular  half-lifes at the r-process
ladders as well as the initial and final state energies. A complete knowledge of the
spin-isospin response of a broad range of nuclei to a various probes is crucial for
not only for the r-process nucleosynthesis, but also for many other applications.
Recently much progress was made in understanding the nuclear spin-isospin response,
both experimentally and theoretically. On the experimental side the matrix elements
of the Gamow-Teller operator $\vec{\sigma} \> \vec{\tau}$ between the initial and
final states were successfully measured using inverse kinematics
\cite{Thies:2012xc}. 
A major theoretical development is the proper inclusion of the tensor force in the
shell model Hamiltonians. 
Monopole component of the nucleon-nucleon force is the same in nuclear medium and
free space, however the monopole effect of the tensor force alters the shell
structure in a significant way \cite{Otsuka:2005zz}. This is because the monopole
component of the tensor interaction changes depending on whether the nucleon spin is
parallel or antiparallel to its orbital angular momentum.  
In most cases the monopole component is an average over all possible spin
orientations, so the tensor component does not contribute for the filled orbits.
However, near the Fermi surfaces where the spin-orbit force splits the orbits, the
$j=\ell +s$ orbit fills first altering the mean field. Indeed residual effective
force between the valence nucleons, beyond that represented by the mean field, is
very well described by the tensor force \cite{Otsuka:2009qs}. A new p-sd shell model
Hamiltonian including up to 2-3 $\hbar \Omega$ excitations can describe the magnetic
moments and Gamow-Teller (GT) transitions in p-shell nuclei well with a small
quenching for the spin g-factor and the axial-vector coupling constant
\cite{Suzuki:2003fw}. These new developments lead to a description of the spin
properties of such nuclei better than the conventional shell model Hamiltonians
resulting in a better description of the weak interactions for astrophysical
applications.  
For example, this new Hamiltonian significantly improves the description of the
cross sections for the reactions $\nu_e +^{12}$C 
\cite{Suzuki:2006qd} and $\nu_e +^{13}$C \cite{Suzuki:2012aa}, potentially important
for scintillator-based neutrino experiments. 
These new Hamiltonians can successfully describe not only the Gamow-Teller, but also
the first-forbidden transitions. Inclusion of the latter terms can significantly
change the lifetimes \cite{Suzuki:2011bk}. These recent advances in the shell model
studies of stable and exotic rare nuclei as well as their use in the description of
the spin-dependent nuclear weak processes are reviewed in references
\cite{Suzuki:2007zza} and \cite{Suzuki:2011zzb}. 

Neutrinos form the bridge between many astrophysical phenomena and laboratory
nuclear physics, investigating stable as well as unstable, exotic, and rare
elements. Exploiting this connection in an intellectually beneficial way
necessitates a multitude of experimental and theoretical efforts.  On the
theoretical side, these include a better understanding of neutrino properties and
improving our knowledge of nuclear structure to calculate neutrino interactions with
nuclei more accurately. On the experimental side, these include measurements of the
spin-isospin response of both stable and exotic nuclei to various probes as well as
measuring salient properties of such nuclei. On the observational side, these
include better determinations of cosmic abundances. 
One of the prizes of this quest is understanding the origin of the elements. At
least to some of us, another prize will be the full appreciation of the considerable
role neutrinos play in the cosmos.

\ack{I was lucky to be associated with Taka Otsuka during the span of both of our
careers and I am  grateful to him for many physics discussions over many years. I
also would like to thank T. Kajino, Y. Pehlivan and T. Suzuki with whom most of the
work presented here was carried out. This work was supported in part 
by the U.S. National Science Foundation Grant No.  PHY-1205024, in part by the
University of Wisconsin Research Committee with funds
granted by the Wisconsin Alumni Research Foundation and in part  through JUSTIPEN
(Japan-U.S. Theory Institute for Physics with Exotic Nuclei) under grant number
DEFG02- 06ER41407 (U. Tennessee).}

\end{document}